\documentclass[conference]{IEEEtran}

  	\usepackage[pdftex]{graphicx}
  	\graphicspath{{../pdf/}{../jpeg/}}
	\DeclareGraphicsExtensions{.pdf,.jpeg,.png}

	\usepackage[cmex10]{amsmath}
	\usepackage{mathabx}
	\usepackage{algorithmic}
	\usepackage{array}
	\usepackage{mdwmath}
	\usepackage{mdwtab}
	\usepackage{eqparbox}
    \usepackage{authblk}
	\usepackage{url}
    \usepackage{xcolor}
    \usepackage[dvipsnames]{xcolor}
\title{Effects of Josephson Junction Non-idealities on Adiabatic Quantum Flux Parametron Circuits}

\author[1,$\dagger$]{Daryoush Shiri}
\author[1]{Likai Yang}
\author[2]{Mohamed A. Hassan}
\author[1]{Philip Krantz}
\author[1]{Eric T. Holland}

\affil[1]{Quantum Engineering Solutions (QES), Keysight Technologies Inc.,
Santa Rosa, CA 95403, USA}
\affil[2]{Design Engineering Software (DES), Keysight Technologies Inc., 
Santa Rosa, CA 95403, USA}
\affil[$\dagger$]{daryoush.shiri@keysight.com}

\begin{document}
\maketitle

\begin{abstract}
Adiabatic quantum flux parametron (AQFP) gate is a promising approach to scale up the cryogenic microwave electronics for superconducting qubit multiplexed control. However, the performance of these circuits depends on the quality of the Josephson junctions which are ideally superconductor-insulator-superconductor (SIS) type following the ideal sinusoidal relation between current and quantum phase. We demonstrate how the non-sinusoidal current-phase relation in Superconductor-Normal metal-Superconductor (SNS) and weak link (WL) junctions affects the speed, delay, and margin of the AQFP gates. The JJ models are defined in the Keysight ADS simulator using symbolically defined device (SDD) method.
\end{abstract}

\IEEEoverridecommandlockouts

\IEEEpeerreviewmaketitle

\section{Introduction}

Adiabatic quantum flux parametron (AQFP) logic gates promise low-dissipation switching and high-speed operation for cryogenic digital and microwave signal processing \cite{Takeuchi_Rev}. \textit{e.g.}, shift registers, analog-to-digital (ADC) converters, decoders, random number generators \cite{RandomNG}, binary neural networks \cite{BNN}. The operation principle of adiabatic gates dates back to R. Landauer in 1960's when he proposed that the energy dissipation in an adiabatic transition of a single-well to double-well potential is as low as $k_BT\ln2$ \cite{Landauer}, which represents the thermodynamic cost of the information loss. Readers are referred to \cite{Takeuchi_Rev} for a detailed review of the field and its later developments. The same concept was also used in Josephson parametric oscillators (JPO) \cite{Bhai_JPOsc}\cite{PK_JPO} and it was shown that high-fidelity qubit readout using two states of a JPO is possible. 
Most recently, AIST and other research centers in Japan have demonstrated the potential of AQFP gates as cryogenic microwave mixers and their applications in the scalable control of superconducting qubits. This approach promised co-integration of qubits and control circuits which significantly reduces the room temperature hardware and cabling bottleneck. 
In this article, we demonstrate hierarchical design and simulation capability of AQFP digital logic gates in microwave frequencies using Keysight ADS simulator. We show how different Josephson junctions can be modeled mathematically without any approximations. The effects of Josephson junction non-idealities (due to different tunneling barriers) on the switching speed and current swing of the AQFP gates are also investigated.
\section{Adiabatic Quantum Flux Parametron (AQFP)}

\begin{figure}[t]
    \centering
    \centering
    \includegraphics[width=75mm]{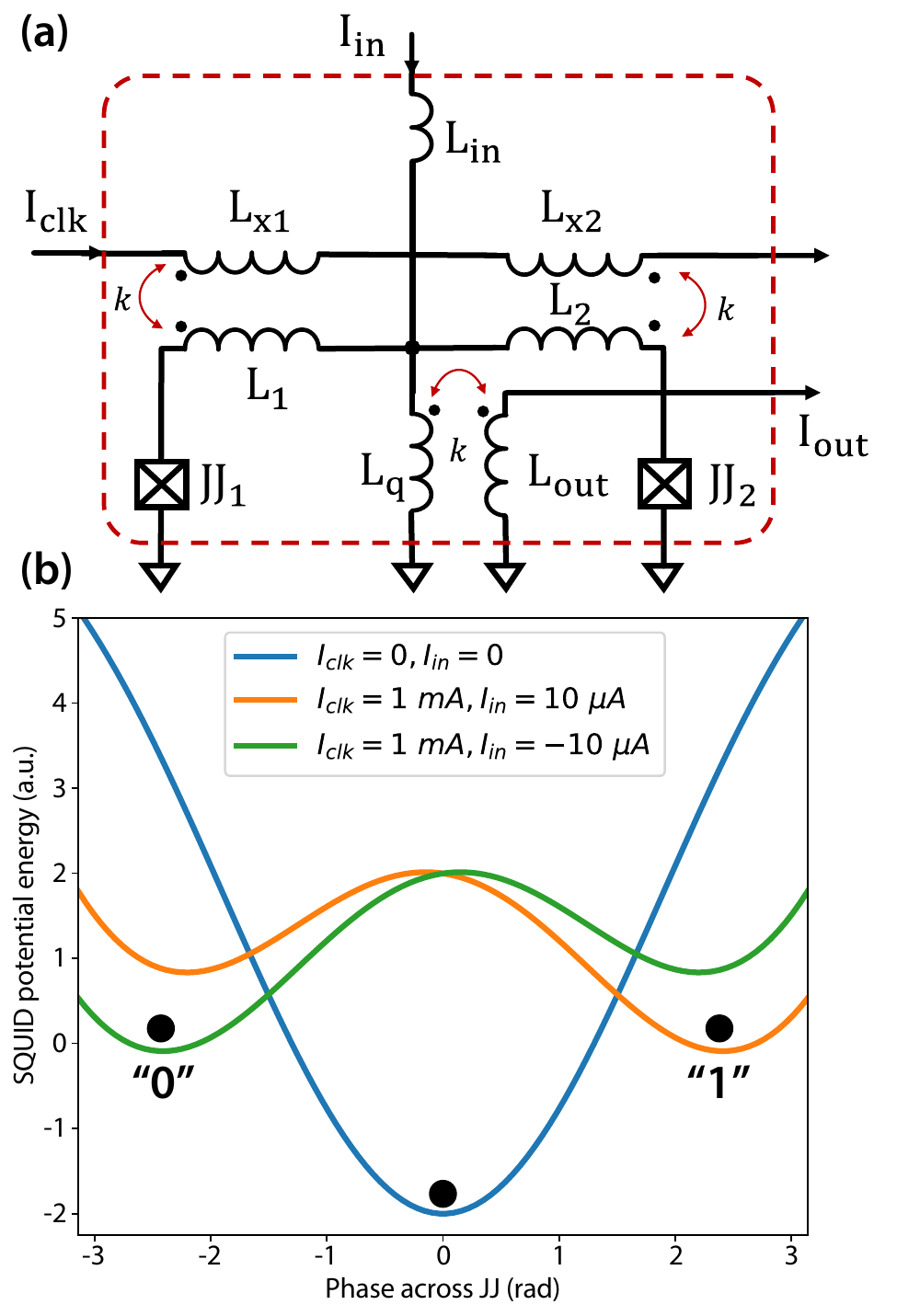}
    \caption{ (a) AQFP buffer, (b) The potential energy in the SQUID loop (normalized to Josephson energy $E_J=I_c\Phi_o/2\pi$ ) as a function of $\phi_x$ which is average of phases across two JJs. Adiabatic transition of potential from single-well to double-well as I$_{clk}$ increases. The final state (clockwise and counter clockwise current in the SQUID) depends on the input current polarity, I$_{in}$. JJs are modeled as shown in Fig. \ref{jjmodel}.}
    \label{aqfp_gate_basics}
\end{figure}

The circuit schematic of a AQFP buffer gate is shown in Fig. \ref{aqfp_gate_basics} (a) which is the superconducting version of Goto parametron \cite{Goto}. The main part of the gate is a SQUID loop made of two Josephson junctions (JJ$_1$ and JJ$_2$) and loop inductances (L$_1$ and L$_2$). The potential energy of the circuit is a function of phases across each JJ which is determined by the magnetic flux induced in the loop by external currents. When the external flux is zero the potential as a function of $\phi_{x}=(\phi_{JJ1}+\phi_{JJ2})/2$ has one minimum at the bottom (See Fig. \ref{aqfp_gate_basics} (b)). By applying the external flux using I$_{clk}$, the single-well potential transits into the double-well potential. The polarity of the input current, I$_{in}$, affects the shape of this gradual (adiabatic) transition and determines the minimums (0 or 1) where the final state falls \cite{Takeuchi_2013}. The clock current is inductively coupled to SQUID loop via L$_{x1}$ and L$_{x2}$ with strength $k$. The output current (state of the buffer) is obtained by coupling L$_{out}$ to L$_{q}$ in Fig. \ref{aqfp_gate_basics} (a). Before discussing the design values and simulation of this gate, let's see how JJs which are the innermost sub-circuits of AQFP gates, are modeled.
\subsection{Josephson Junction Model as a Sub-circuit}

\begin{figure}[t]
\centering
\includegraphics[width=75mm]{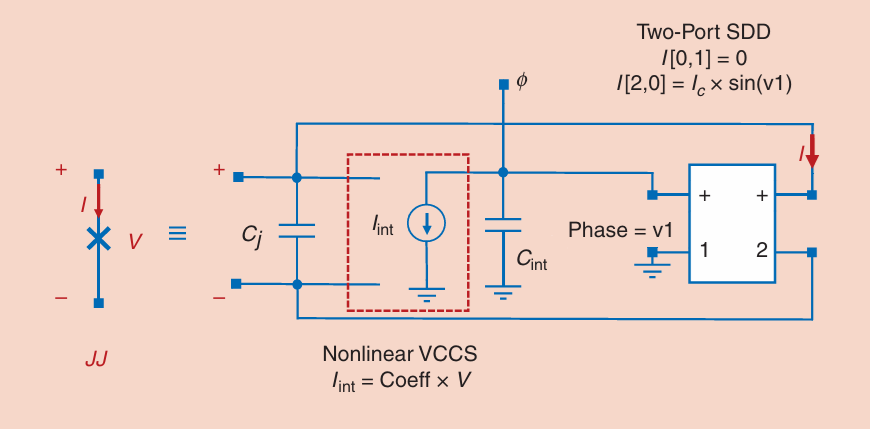}
\caption{The circuit symbol and the mathematical modeling of Josephson junction using a symbolically defined device (SDD). The sinusoidal term, $\text{sin}(v_1)$, can be replaced by a polynomial or a more sophisticated function of $\phi$ for non-ideal JJs. \textbf{Coeff} is $2\pi C_{int}/\Phi_o$. This figure is adapted from \cite{Shiri2024}.}
\label{jjmodel}
\vspace{-\baselineskip}
\end{figure}

A Josephson junction with ideal insulating tunneling barrier, is known as superconductor-insulator-superconductor (SIS) junction, has a sinusoidal relationship between the current and the quantum mechanical phase (\textit{i.e.}, phase change of Cooper pair wave function across the SIS junction) \cite{Josephson1962}. The Josephson equations of current and voltage are;
\begin{equation}
I=I_c\text{sin}(\phi)~~\text{and}~~\phi = \frac{2\pi}{\Phi_o} \int V dt,
\label{phase}
\end{equation}
where $I_c$ is the junction critical current and $\Phi_o=2.07\times10^{-15}$\,Wb is the magnetic flux quantum.
Eq. (\ref{phase}) is modeled in Keysight ADS simulator using the symbolically defined device (SDD) without incurring any approximation \cite{Shiri2024}. Fig. \ref{jjmodel} shows the model which is used as a cub-circuit for the rest of this study. The two port SDD converts the phase $\phi$ into current according to Eq. (\ref{phase}). The quantum phase $\phi$ is created by integrating the voltage using an integrator made of a capacitor, C$_{int}$, and a current source, I$_{int}$ (see Fig. \ref{jjmodel}).  
\subsection{Non-sinusoidal Current-Phase Relationship}

\begin{figure}[t]
\centering
\includegraphics[width=75mm]{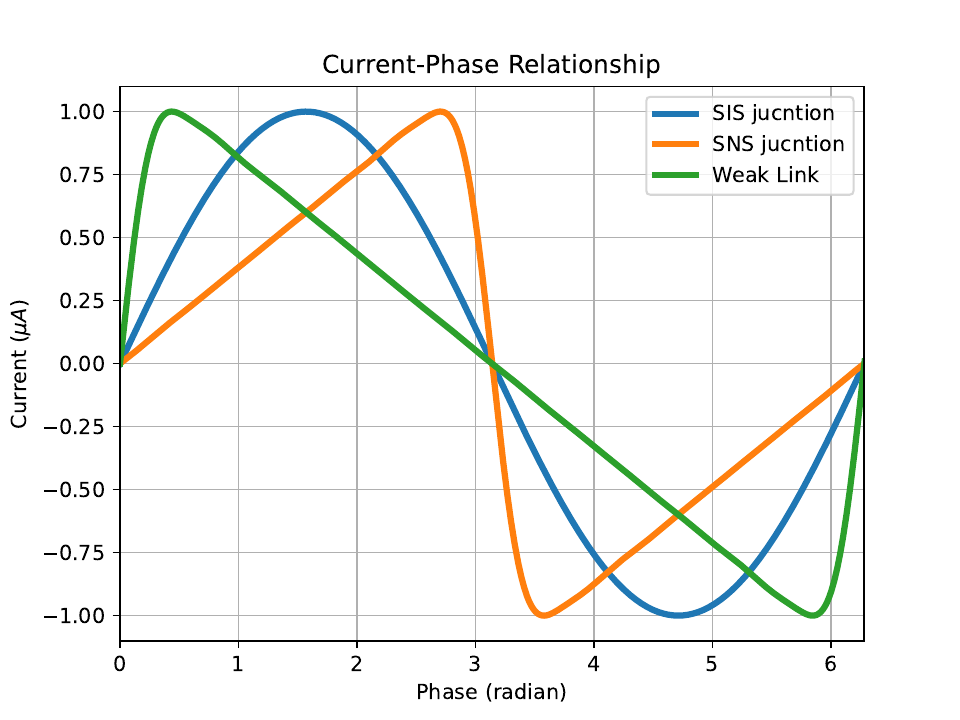}
\caption{Current-phase relationship of three different Josephson junctions modeled in Kesyight ADS using SDD block. The expansion coefficients of odd harmonics for SNS and WL in Eq. (\ref{nonsin}) are positive and negative, respectively.}
\label{3jjmodels}
\vspace{-\baselineskip}
\end{figure}

Depending on the type of the tunneling barrier in the Josephson junction, the current-phase relationship can deviate from that of SIS \cite{Thompson2023}. For example, if two superconductors are connected by a narrow superconducting metal, called constriction or weak link (WL), the current-phase relationship is a skewed sinusoidal or triangular. Also, a junction with a normal metal layer between superconducting layers (SNS) deviates from the $\sin(\phi)$ relation of Eq. (\ref{phase}). The SDD block facilitates modeling of these types of junctions by allowing $\sin(\phi)$ to be replaced by polynomials or Fourier expansion of $I-\phi$ characteristics found in theory, measurement, or in general by solving Usadel equation in a superconducting multi-layer device \cite{Colletta_Usadel}. The polynomial current-phase model for SNS and WL can be written as: 
\begin{equation}
I/I_c = \sum_{n=1}^{N} \alpha_n \text{sin}(n\phi),
\label{nonsin}
\end{equation}
where the values of $\alpha_n$ and N depend on the type of JJ. For SIS, only $\alpha_1=1$ since N~=~1. In this article, we use the polynomial models of SNS and WL and $\alpha_n$ values proposed by \cite{Thompson2023} to demonstrate the effect of these junctions on the performance of an AQFP buffer (or NOT) gate. The difference between the current-phase relation in Eq. (\ref{nonsin}) for SNS and WL lies in the opposite sign of odd harmonics. The $I-\phi$ relationship of three JJ models are shown in Fig. \ref{3jjmodels}.

\section{AQFP Buffer}
\begin{figure}[t]
\centering
\includegraphics[width=75mm]{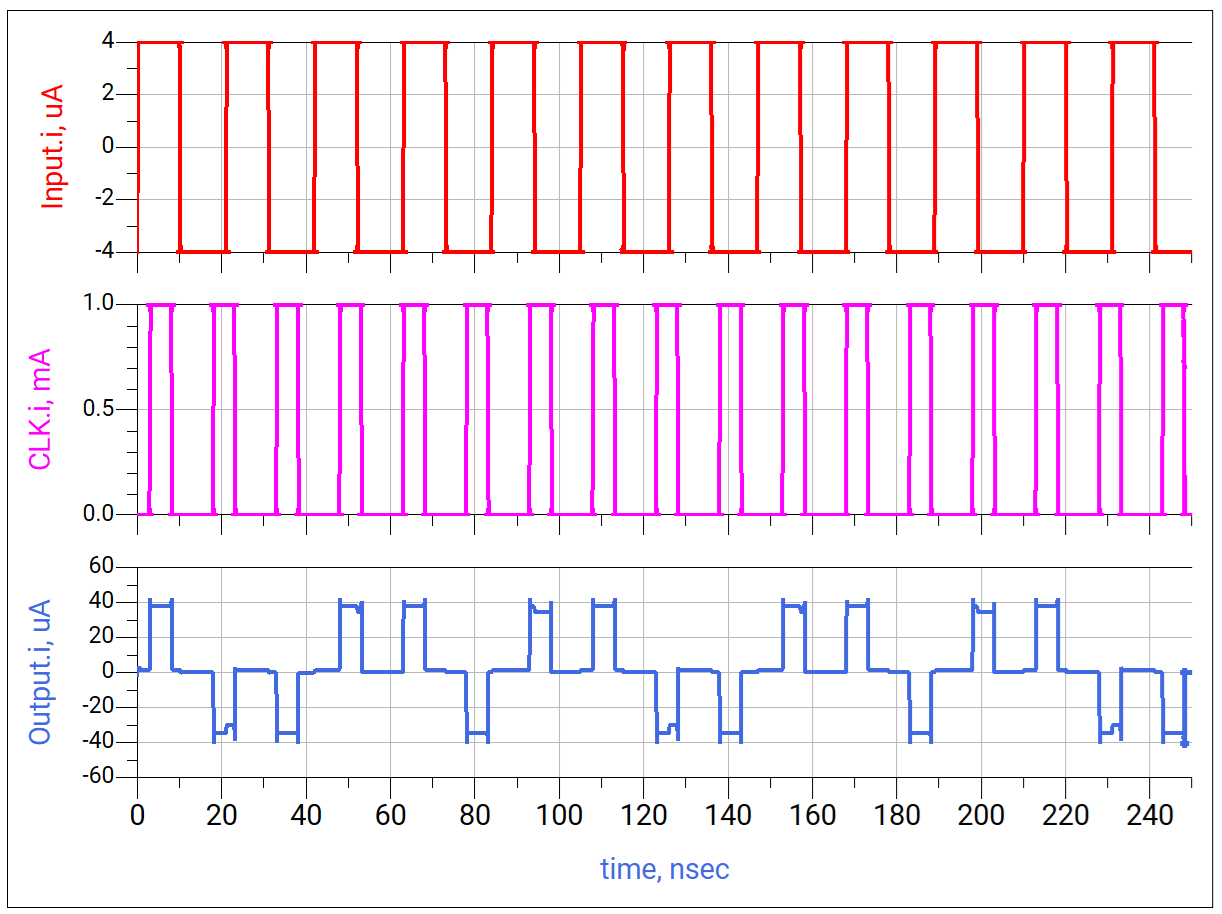}
\caption{AQFP buffer simulation results in Keysight ADS simulator. The clock (magenta) samples the input current (red) on the rising edges and copies a replica of it with one clock pulse duration at the output (blue).}
\label{buffer}
\vspace{-\baselineskip}
\end{figure}

\begin{figure}[t]
\centering
\includegraphics[width=75mm]{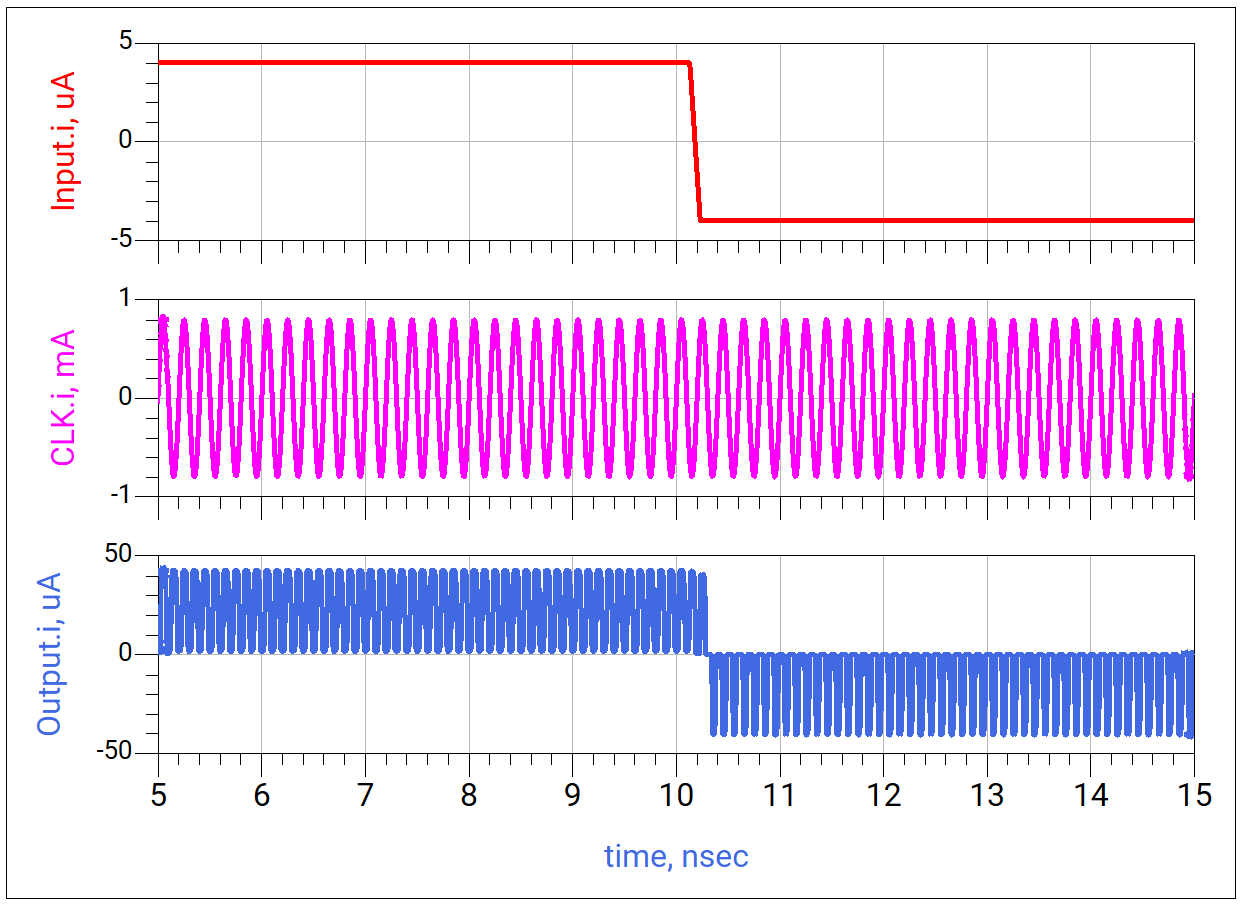}
\caption{AQFP buffer simulation with sinusoidal 5~GHz clock. The red, magenta and blue plots are input current, clock, and the output current, respectively. The polarity of the input current determines the polarity of output samples.}
\label{buffer_sine}
\vspace{-\baselineskip}
\end{figure}

The role of this buffer is to sample the input signal at the rising edge of the clock while keeping one sample per clock cycle. The act of sampling occurs at the time instant when the clock current has reached half a quantum of flux ($\Phi_o/2$) in the SQUID loop. The circuit schematic of AQFP buffer is shown in Fig. \ref{aqfp_gate_basics} (a). The polarity of the output current depends on the polarity of the sampled input current. 
The values of JJ parameters and inductances are adapted from \cite{Takeuchi_2013} which are based on the AIST planarazied Niobium process used for cryogenic microwave circuits. The JJ critical currents are $I_{c1}=I_{c2}=50$\,{\textmu}A. The Stewart-McCumber parameter, $\beta_c < 1$, is chosen to keep the junctions in the overdamped regime for fast switching, suitable for clocks rates $>$ 5\,GHz. The mutual coupling coefficients determine the onset of half-quantum flux switching and are found from $M=k\sqrt{L_{x1}L_1}$. The inductance values of the AQFP buffer gate are $L_{x1} = L_{x2} = L_1 = L_2 = 2.64$\,pH, $L_{in} = 15$\,pH, $L_{q} = 10.5$\,pH, and $L_{out}= 10$\,pH and $0.3< k <0.5$ for all inductive couplings. Fig. \ref{buffer} shows the transient-time simulation of AQFP buffer operation. On the rising edge of the clock (magenta), samples of the input current (red) are put at the output (blue). The rise time and fall time of the clock in the simulation is 0.1\,ns and the peak current is above 800\,{\textmu}A. Unlike classical logic circuits, the clock can be a continuous microwave signal, however design parameters must be optimized for high frequency operation \textit{e.g.}, design of impedance matched lines and post-layout simulation of cross talk \cite{Takeuchi_2017}. 
Fig. \ref{buffer_sine} shows the operation of the same AQFP buffer with 5~GHz clock. As can be seen, the high frequency samples of the input current (red) are created at the output (blue). By inverting the mutual inductance sign of the $L_q$ and $L_{out}$ the AQFP buffer works as a NOT gate. More importantly, this cell can be used as a mixer considering the input current as the base-band modulating waveform and the clock as the RF signal to be modulated. The AQFP buffer (or NOT) gate is the building block of modulated carrier generator for control (rotation) and frequency tuning of superconducting quantum bits (qubits). This circuit technology promises high integration, scalability, and flexibility by providing frequency multiplexing as proposed in \cite{Mukai_2026}.
\section{AQFP Shift Register}
\begin{figure}[t]
\centering
\includegraphics[width=75mm]{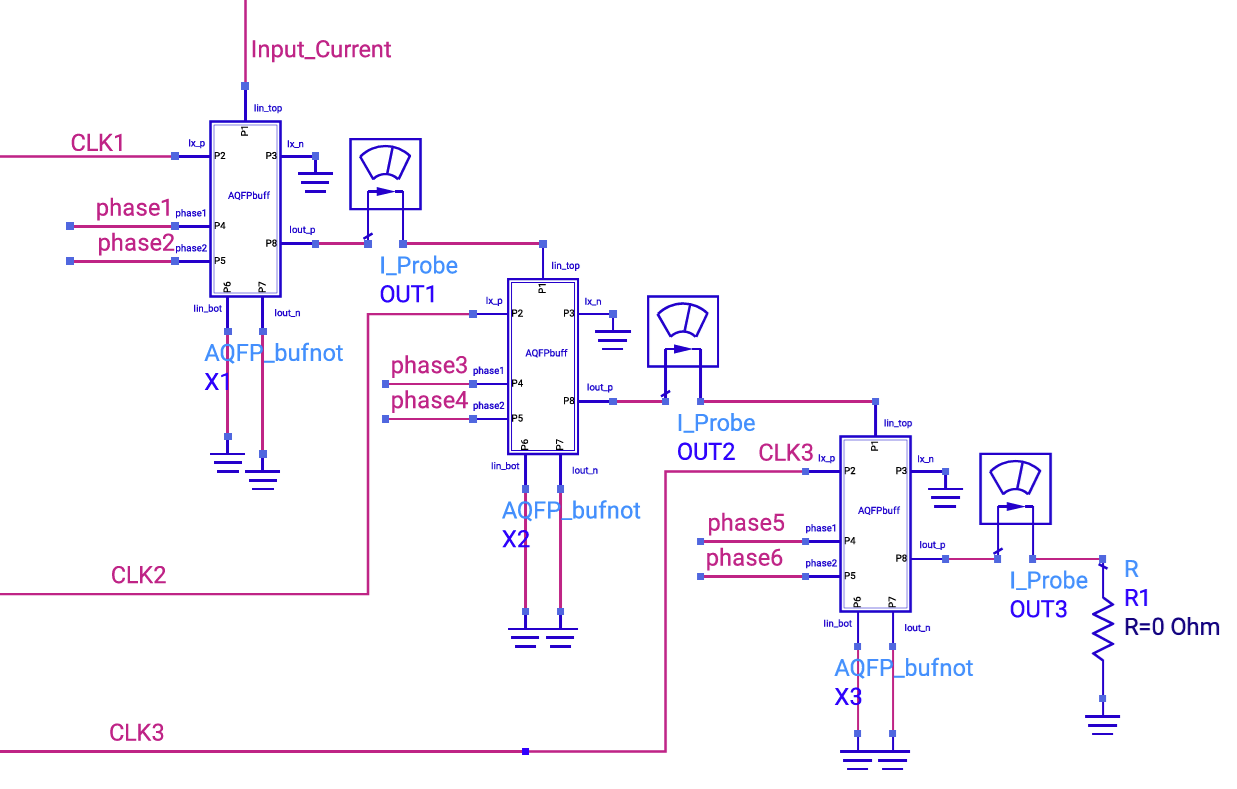}
\caption{A 3-stage AQFP shift register. The input current is fed from the top node of the left-most AQFP gate. The output current of each stage is cascaded to the next stage as the input. The clock of each stage is phase-shifted by 90 degree or 1/4'th of a period relative of the preceding clock.}
\label{3stage}
\vspace{-\baselineskip}
\end{figure}

\begin{figure}[t]
\centering
\includegraphics[width=75mm]{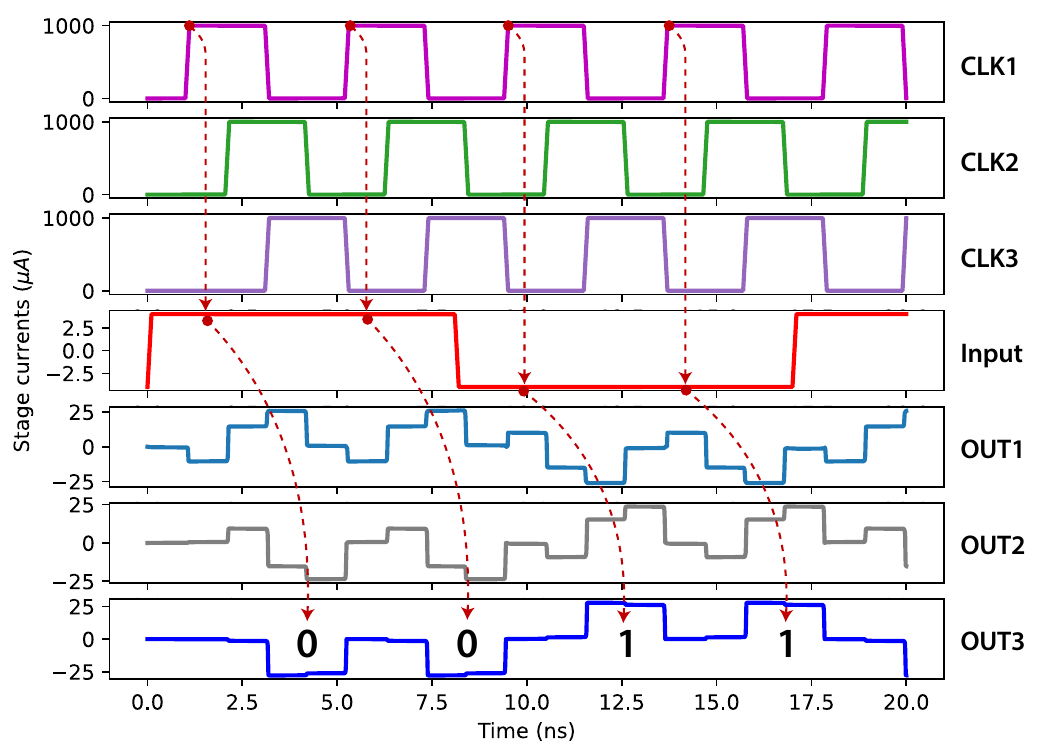}
\caption{Transient-time simulation of a 3-stage AQFP shift register. The samples which are taken from the input current (red) and inverted, appear after one clock period at the output (blue).}
\label{3stage_waves}
\vspace{-\baselineskip}
\end{figure}

Ease and precision of JJ-based circuit modeling in Keysight ADS allows hierarchical design of highly-integrated AQFP gates and rapid single flux quantum (RSFQ) gates alike. Here we demonstrate the design of a 3-stage AQFP shift register made of 3 buffers (or NOTs). The clock for each AQFP buffer is a 90 degree or 1/4\textsuperscript{th} of a period phase-shifted version of the clock in the previous gate. The input current is entering the first stage from the above as shown in Fig. \ref{3stage}. The input current of the 2\textsuperscript{nd} and the 3\textsuperscript{rd} stage are fed from the output currents of the 1\textsuperscript{st} and the 2\textsuperscript{nd} stage, respectively in a cascade topology. The transient-time simulation results are shown in Fig. \ref{3stage_waves}. The three phase-shifted replicas of clock are shown at the top. On the rising edge of the first clock (CLK1) a sample of the input current (red) is created and these inverted samples (due to NOT action) appear after one complete period of the clock at the output (blue) \textit{i.e.}, at the rising edge of CLK3. This demonstrates the feasibility of using AQFP gates in multi-GHz digital data processing at cryogenic temperatures.
\section{Effect of non-sinusoidal $I(\phi)$ on AQFP gate}
\begin{figure}[t]
\centering
\includegraphics[width=75mm]{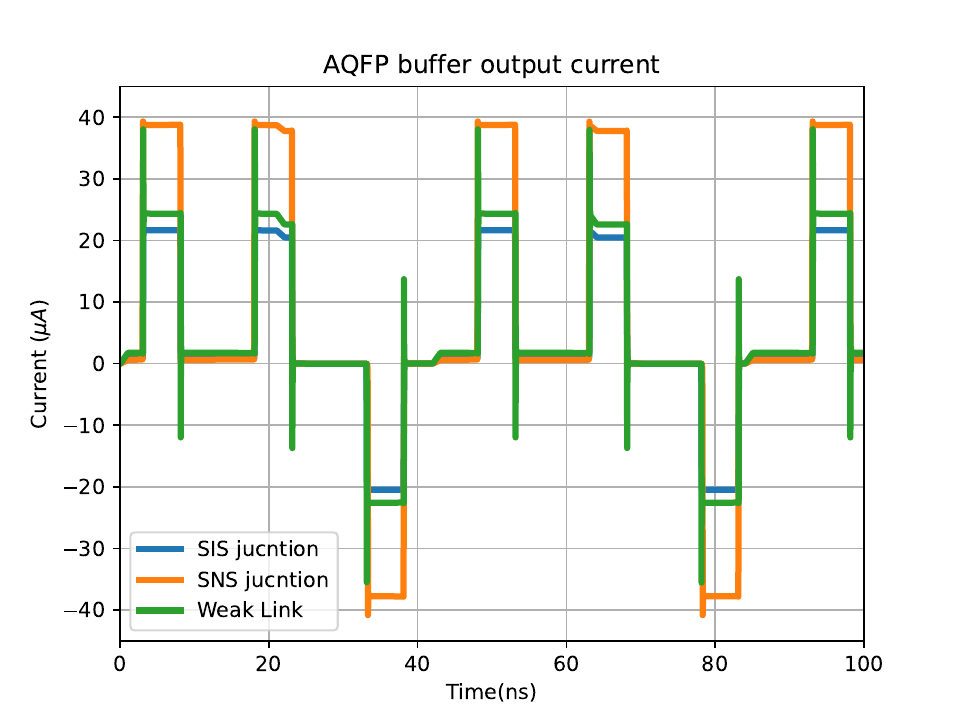}
\caption{The output current of a AQFP buffer when the JJs are of SIS, SNS, and WL type. The input current, clock timings, and amplitudes are the same as the simulations of Fig. \ref{buffer}.}
\label{3jj_effect1}
\vspace{-\baselineskip}
\end{figure}

\begin{figure}[t]
\centering
\includegraphics[width=75mm]{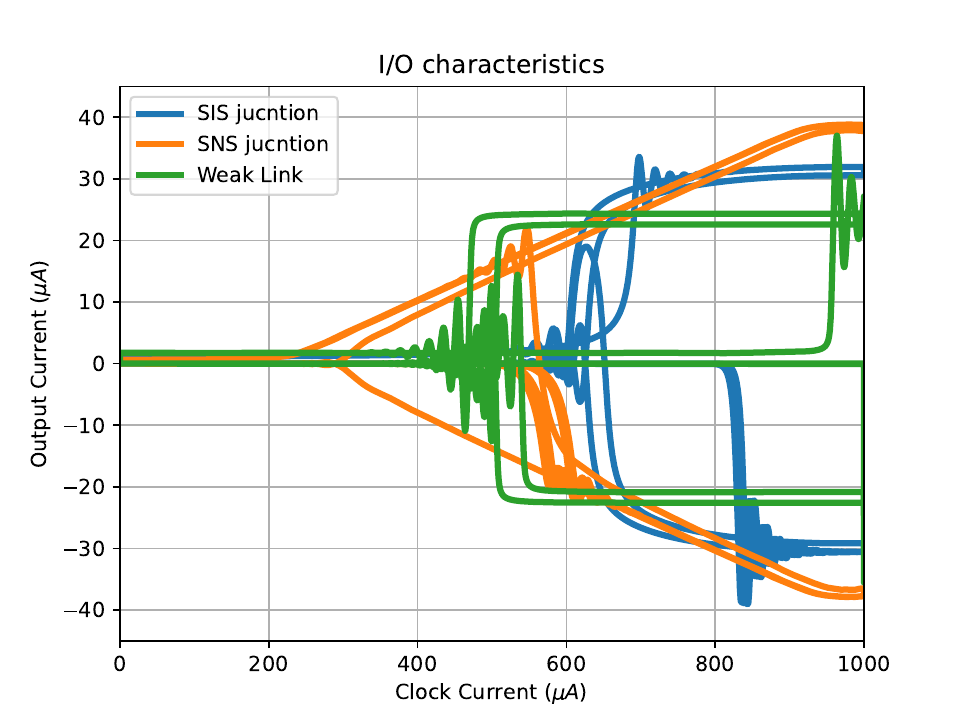}
\caption{The output current of a AQFP buffer versus the clock current showing the onset of switching and current swing for three different Josephson junctions. The input current, clock timings, and amplitudes are the same as the simulations of Fig. \ref{buffer}.}
\label{IO_plot}
\vspace{-\baselineskip}
\end{figure}

In this section, we demonstrate how the quality of Josephson junction tunneling layer (insulator, weak link, and normal metal) affects the performance of the AQFP buffer gate. As mentioned before, the general current-phase relationship for SIS, SNS and WL junctions are modeled as a sub-circuit and used in higher hierarchies of AQFP designs. The current-phase relationships of these three types of Josephson junctions were presented in Fig. \ref{3jjmodels}.
Fig. \ref{3jj_effect1} shows the output current of the AQFP buffer when the JJs are of SIS, SNS and WL type. In contrast to the SIS case when the peak-to-peak of the output current was $\pm$20\,{\textmu}A, this swing is almost doubled for SNS \textit{i.e.}, $\pm$40\,{\textmu}A, and for WL it is around $\pm$25\,{\textmu}A. To put this in a better perspective, we look at the I/O characteristics plots of the AQFP buffer for three types of junctions (see Fig. \ref{IO_plot}). This plot shows the output current versus the clock current. The speed of the gate, transition margins, and switching delay can be inferred from this eye-diagram-like plots. The SNS junction evidently causes more current swing at the output but the switching is not as steep as SIS and WL junctions. That can be explained by the slower increase of current versus phase shown in Fig. \ref{3jjmodels}. It can be said that, for the same amount of flux or phase required for switching, the SNS is the slowest one. On the other hand, WL has a sharper increase of current with phase and the clearer margin is obtained with smallest amount of current among the three types of junctions. In conclusion, SIS is evidently lies between the two extremes (SNS and WL) due to its symmetric (sinusoidal) current-phase relationship. 

\section{Conclusions}
We demonstrated a hierarchical and scalable approach for modeling and simulation of cryogenic digital AQFP gates in microwave regime using Keysight ADS. This approach enables a unified device-to-circuit modeling workflow critical for scalable cryogenic control co-design. We showed how different types of JJs can be modeled using symbolically defined devices (SDD) and investigated their effects on the performance of AQFP buffer. Josephson junction non-idealities reshape the potential landscape of AQFP gates, causing deviations from adiabatic switching and introducing excess dissipation above the Landauer limit, while the specific form of $I-\phi$ determines how this thermodynamic penalty manifests through trade-offs in switching dynamics, timing delay, and stability margins. 

\bibliographystyle{IEEEtran}
\bibliography{IEEEabrv,Maindraft_APMC2026_DShirietal}
\end{document}